\documentstyle[aps,multicol]{revtex}
\input epsf
\title
{
Nonuniqueness and Turbulence
}
\author{Mark A. Peterson}
\address{
Physics Department, Mount Holyoke College, South Hadley MA 01075 USA
}
\date{October 9, 1997}
\begin{document}
\maketitle
\begin{abstract}
The possibility is considered that turbulence is described by
differential equations for which uniqueness fails maximally, 
at least in some limit. 
The inviscid Burgers equation,
in the context of Onsager's suggestion that turbulence should
be described by a negative absolute temperature,
is such a limit.  In this picture, the onset of
turbulence coincides with 
the proliferation of singularities which characterizes 
the failure of uniqueness.
\end{abstract}
\begin{multicols}{2}
\section{Nonuniqueness}
The existence and uniqueness
of solutions to the differential equations of physics
is seldom an issue.
The very fact that these equations describe physical reality seems to
argue that their
solutions must exist and be unique.  Textbook examples of
nonuniqueness, for example Clairaut's equation (Ref. 
\cite{courant}, p. 94), seem like
exceptional cases, which would not arise in physics in any case.

There is, however, a system of
differential equations, arising as a limiting case of
a much studied physical problem, in which uniqueness fails for
{\em every} solution
at {\em every} time.
The system is singular Laplacian growth \cite{peterson1}, \cite{peterson2}.
One might call this behavior ``maximally nonunique.''  It is so
different from the usual behavior of differential equations that 
it hardly seems like a differential system at all.
The way it occurs is the following:  the theory of singular
Laplacian growth describes the motions of certain singularities
of a conformal map, which are all located on the unit circle
in the complex plane.  They move on the circle, but they can
also ``split,'' introducing new singularities, at any time.
This behavior, which does not sound very remarkable, 
essentially
implies the maximal nonuniqueness property.
In Ref. \cite{peterson2} the term ``fragile'' was suggested 
for such a system,
because its distinguishing feature is that its 
singularities can break apart.

It is natural to ask if other physical systems might
reduce to a fragile system in some limit, and if the fragile
property manifests itself in the behavior of the system.
Turbulence is certainly a candidate to be a
fragile system:  like Laplacian growth, turbulence is 
a phenomenon in which a differential equation has
unexpectedly complex solutions.  In this paper I argue
that turbulence is a system that has a fragile (maximally
nonunique) limiting case.  The limiting case is the 
Burgers equation.
(On rather different grounds a resemblance between the problems of turbulence
and Laplacian growth has been
noted by Hastings and Levitov \cite{Hastings}).

\section{Inviscid Burgers Equation}
The Burgers equation \cite{Burgers}, \cite{strang}
\begin{equation}
\frac{\partial u}{\partial t}+u\frac{\partial u}{\partial x}=\nu \frac{\partial^2 u}{\partial x^2}
\end{equation}
has played a shifting role in the problem of
turbulence.  It was first introduced as a one-dimensional zero pressure version of the
Navier-Stokes equation (with $\nu$ the viscosity)
in the hope that it might exhibit in its solutions some of the 
complexity of turbulence.  By the Hopf-Cole trick, though, it was found to be
equivalent to a linear diffusion equation, so that the initial value problem 
could be solved explicitly:  one can write a formula for the solution!  That
is surely too simple to be turbulent.  Attention was thus focussed 
 on the only case which still seemed promising, the $\nu\rightarrow 0$
limit, the inviscid Burgers equation, which is a conservation law for $u$.
In this limit the solution may develop discontinuities (``shocks'').  
The distribution of shocks, and their development in time, for random initial data,
is a problem called Burgers turbulence.

The reason that one should think of this as a $\nu\rightarrow 0$ limit, and
not simply as a $\nu=0$ version of the equation, is that for $\nu=0$ the
solution is ambiguous.  To be sure, 
the solution of the inviscid equation is immediate
by the method of characteristics:  $u$ is constant along lines $dx/dt=u$
(hence straight lines in the x-t plane).
But if the characteristics cross, which they certainly will do in the case of
random initial data $u(x,0)$, then one has multiple 
determinations of $u(x,t)$,
as in Fig. 1.  The resolution of this difficulty is  
to take the formula for  $u(x,t)$ in case $\nu>0$, which is unambiguous,
and let $\nu$ approach zero.  The result is as shown in Fig. 2:  a discontinuity 
forms at a definite location, separating one determination of $u$ from another.  
The same procedure shows that shocks interact in a definite manner, coalescing
as shown in Fig. 3 when one overtakes or collides with another.    

The rules for how these discontinuities form and interact were known long
ago in the theory of compressible gas dynamics, where they are called
the Rankine-Hugoniot jump conditions (see Ref. \cite{courant}, pp. 488-90, 
Ref. \cite{strang}, p. 596).  
They are justified by appeal to the
second law of thermodynamics:  while $u$ is conserved, entropy must increase
in the shock.
That is what the argument with $\nu>0$ also does:  we resolve the ambiguity
in the inviscid Burgers equation
by putting in a dissipative term with the right sign.  The second law of
thermodynamics says mechanical
energy should be dissipated in the shock, and not
created.  It is this condition which leads to the interaction rule illustrated
in Fig. 3.

\section{Negative Temperature}
Now recall the suggestion of Onsager, as emphasized by Alexandre Chorin,
among others, that a statistical theory
of the turbulent steady state should be characterized by a negative
temperature (see Ref. \cite{chorin}, chapter 4). 
What is meant is that there is a Maxwell-Boltzmann probability
distribution
\begin{equation}
P\sim e^{-E/kT}
\end{equation}
which, instead of giving more weight to the low energy microstates
in the ensemble description of the macrostate, as is usual, gives
more weight to high energy microstates.  (This statistical
``temperature'' of turbulence has nothing to do with usual 
thermodynamic temperature,
which is only weakly coupled to the mechanical degrees of freedom
of interest.)  Let us accept this idea for the moment.
The second law says that entropy should increase, 
or perhaps better, in this more general situation, that information 
should be lost, in irreversible
processes.  Stated in terms of
the free energy
\begin{equation}
F=E-TS
\end{equation}
(and remembering $T<0$)
this says that free energy should {\em increase} in the approach to 
the steady state.
This is opposite to what is usual.  Putting Onsager's idea together with
the ideas of Burgers turbulence thus requires that mechanical energy
should be created in the shock rather than dissipated.
We see that the limit in the inviscid Burgers equation
should be $\nu\rightarrow 0_-$, i.e., we should
imagine the viscosity approaching zero through {\em negative} values,
from {\em below}.

The inviscid Burger's equation does not contain temperature or $\nu$ explicitly,
of course, but the rules for how the
shocks form and interact are now different.   The following argument
gives the simple idea which is at the heart of this paper.  We do
not try to give it an elaborate justification, since, like many simple ideas,
it may contain some truth even if the arguments are wrong.  The argument
is that there is a symmetry of the (general) Burgers equation,
\begin{equation}
	\nu\rightarrow -\nu, \qquad t\rightarrow -t, \qquad x\rightarrow -x.
\end{equation}
We can convert the $\nu<0$ inviscid Burgers equation into the familiar
$\nu>0$ inviscid Burgers equation by reversing the signs of $x$ and $t$,
i.e., reflecting graphs like Fig. 3 through the origin.  The result is
Fig. 4, in which a single shock has spontaneously split into two,
with no reference to initial conditions.  
This is the fragile property.  If shocks can split at any time, as
indicated, then the solutions to the differential equation are 
maximally nonunique.  

\section{Interpretation}
Our aim in the previous section was to show that the equations which
describe fluid flow, the Navier-Stokes equations, become maximally
nonunique, or ``fragile,'' in some limit.  This limit is rather far
removed from physical reality, however.  One naturally wonders how,
if at all, the nonuniqueness
property might manifest itself in a real system.
The example of Laplacian growth encourages one to think that real
processes would not completely obscure the underlying ``fragile''
processes \cite{peterson2}.

The system we have imagined is characterized by two temperatures:
a ``superhot'' negative temperature, which describes the ensemble
of microscopic entities which make up the turbulent state,
and the usual temperature, which describes the ensemble at the
molecular level.  These two ensembles interact only weakly, it
has been argued, but, again by the second law of thermodynamics,
to the extent that they interact, there must be a flow of energy
from the first to the second, from the turbulent ensemble to the
molecular ensemble.  This picture is reminiscent of the
Kolmogorov cascade idea (Ref. \cite{chorin}, chapter 3), 
and suggests identifying the turbulent
ensemble with the ``energy range,'' (a hypothetical range in $k$ space 
containing most of the energy), and the molecular ensemble with
the ``dissipation range,'' (a range in $k$ space, disjoint from the
energy range, in which dissipation is important), i.e., assigning
a temperature gradient to the Kolmogorov picture, with negative
temperature at small $k$ and positive temperature at large $k$.
It is in the energy range, then, at small $k$, that the fragile
processes of nonuniqueness would occur.

These processes extract mechanical energy from the negative
temperature ``bath,'' and this energy must ultimately derive from the
forces maintaining the turbulence.  Thus the onset of fragile
processes would appear, on the macroscopic scale,
as an increase in resistance to these
applied forces.  At a more microscopic level it would be the
onset of nonuniqueness, allowing splitting and proliferation of
microscopic entities as in Fig. 4.  At a still more microscopic
level, corresponding to the dissipation range, the usual
picture would apply, and one would have dissipative processes
like Fig. 3.  In this view turbulence {\em is} 
the visible manifestation of nonuniqueness on the 
intermediate scale of the
energy range, and the work done by external forces
goes directly into the proliferation of singularities
at that scale, and only indirectly into dissipation.

In terms of modelling turbulence, it suggests that the proliferative
processes of the energy range, which are still hypothetical,
may be like the dissipative processes of the
dissipation range, but time reversed 
(and on a larger length scale).
In this way the abstract picture suggested here might persist
in a more realistic dynamics. 

The issue of nonuniqueness may also be relevant to CFD modelling of
turbulence.  The algorithms of differential equation solvers are not set
up for equations which do not have unique solutions.  Experience with
Laplacian growth confirms that proximity to a nonunique model may
indicate trouble for conventional numerical solutions.

\end{multicols}
\newpage
\begin{figure}[htb]
\centerline
{
\epsfxsize=12cm
\epsfbox{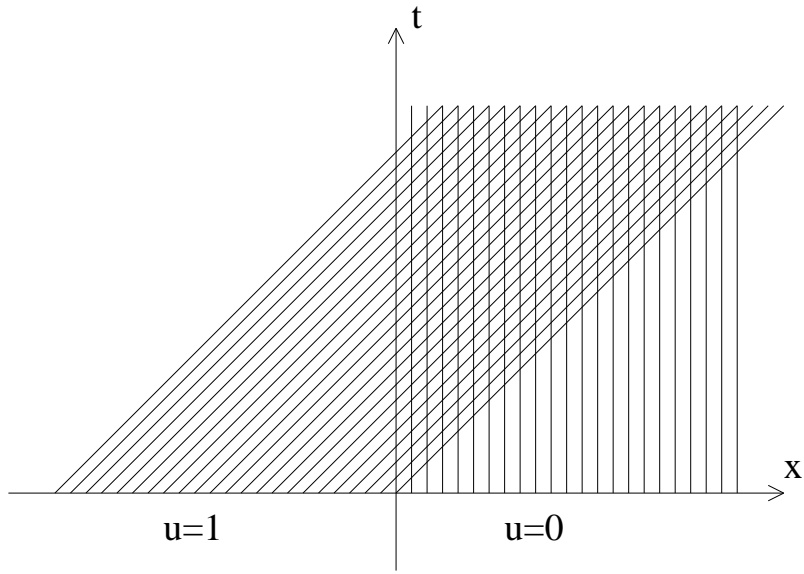}
}
\caption
{
The $\nu=0$ Burgers equation with initial data $u=1$ for $x<0$ and $u=0$ for $x\geq0$
determines u(x,t) to be $1$ where the characteristics have slope $1$ and $0$ where
they are vertical.  In $t\geq x\geq 0$ this determination is ambiguous.
}
\end{figure}
\begin{figure}[htb]
\centerline
{
\epsfxsize=12cm
\epsfbox{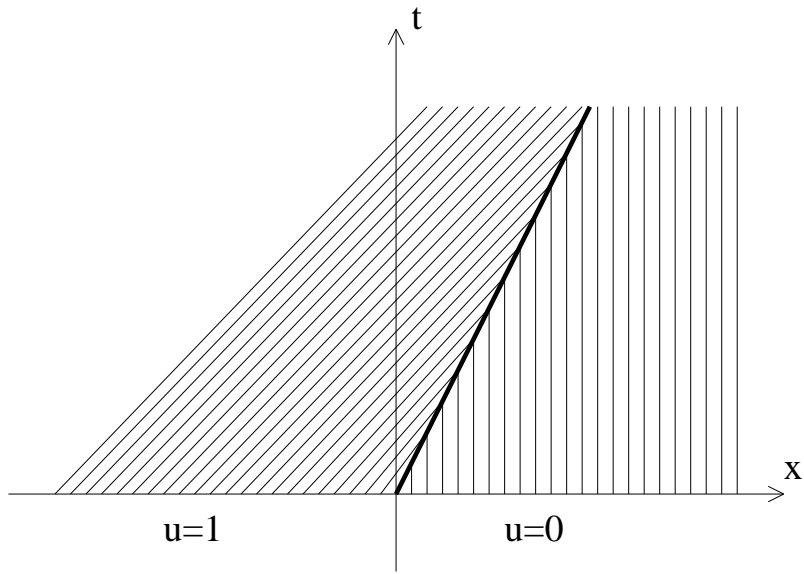}
}
\caption
{
The $\nu\rightarrow 0$ limit of the $\nu>0$ Burgers equation removes the ambiguity
of Fig. 1.  A shock develops between the two regions.
}
\end{figure}
\begin{figure}[htb]
\centerline
{
\epsfxsize=12cm
\epsfbox{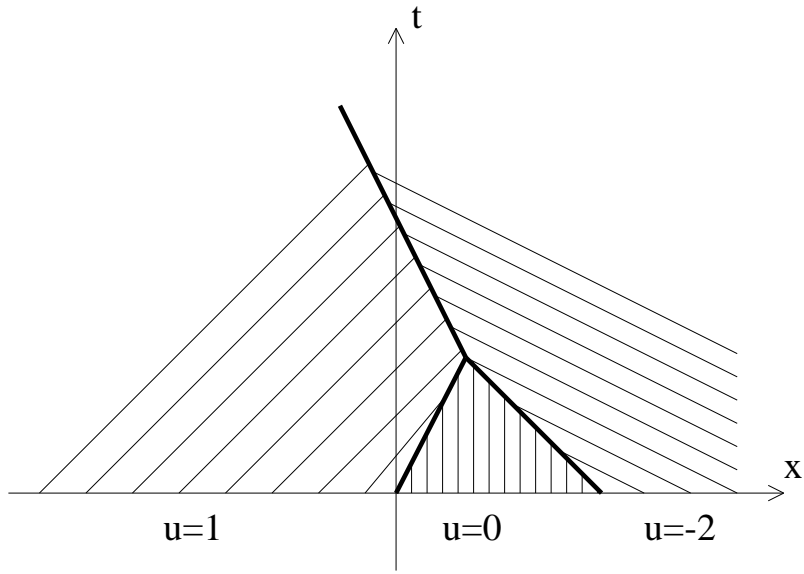}
}
\caption
{
The $\nu\rightarrow 0$ limit of the $\nu>0$ Burgers equation determines how
shocks interact:  they coalesce, with dissipation of energy.
}
\end{figure}
\begin{figure}[htb]
\centerline
{
\epsfxsize=12cm
\epsfbox{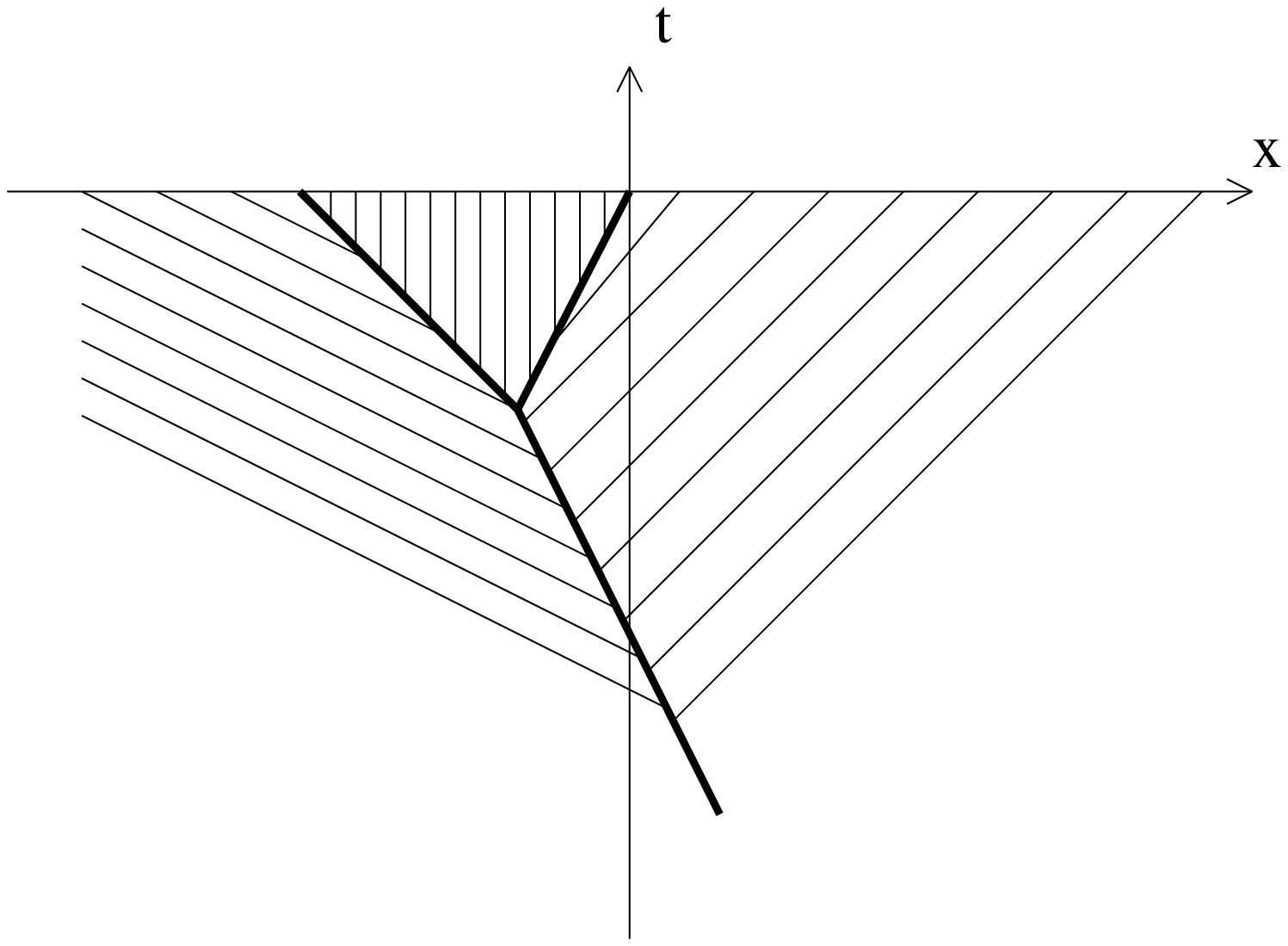}
}
\caption
{
It is suggested that the $\nu\rightarrow 0$ limit of the $\nu<0$ Burgers
equation should be understood as the time reversal of the usual
inviscid Burgers equation.  This is Fig. 3 upside-down.  It shows
a shock splitting at an arbitrary time.  This mechanism,
if it actually occurs, makes the time evolution of the Burgers equation
nonunique at every time, i.e. maximally nonunique.
}
\end{figure}


\begin{thebibliography}{99}
\bibitem{courant} R. Courant and D. Hilbert, {\it  Methods of Mathematical Physics}, vol 2
(Interscience Publishers, 1962).
\bibitem{peterson1}  M.A. Peterson, {\bf Phys. Rev. Lett. 62}, 284 (1989).
\bibitem{peterson2}  M.A. Peterson, cond-mat/9710046.
\bibitem{Hastings} M.B. Hastings and L.S. Levitov, cond-mat/9607021.
\bibitem{Burgers}  J.M. Burgers, {\it The Nonlinear Diffusion Equation},
	(Riedel, Dordrecht, 1974).
\bibitem{strang} G. Strang, {\it Introduction to Applied Mathematics},
(Wellesley-Cambridge Press, Wellesley MA, 1986).
\bibitem{chorin}  A.J. Chorin, {\it Vorticity and Turbulence}, (Springer, 1994).

\end{thebibliography}
\end{document}